\documentclass[aps,prb,notitlepage,superscriptaddress,showkeys,12pt]{revtex4-1} %Which version of RevTex?

\usepackage{amsmath}
\usepackage{amssymb}
\newcommand{\ket}[1]{\ensuremath{\vert#1\rangle}}
\newcommand{\bra}[1]{\ensuremath{\langle #1|}}
\newcommand{\ens}[0]{\ensuremath} %Kurzschreibweise für mathematische Definitionen
\newcommand{\anfEngl}[1]{``#1''} %englische Anführungszeichen
\DeclareMathOperator{\Spur}{Tr} %
\newcommand{\x}[0]{\ens{\otimes}} %Tensorprodukt
\renewcommand{\phi}[0]{\ens{\varphi}} %schöneres Phi
\newcommand{\iE}[0]{\ens{\mathrm{i}}}
\newcommand{\EZ}[0]{\ens{\mathrm{e}}} %Eulersche Zahl
\newcommand{\pr}[1]{\ens{\ket{#1}\bra{#1}}} %Projektion auf #1
\newcommand{\N}[0]{\ens{\mathbb{N}}}
\newcommand{\C}[0]{\ens{\mathbb{C}}}
\newcommand{\cH}[0]{\ens{\mathcal{H}}}
\newcommand{\Mg}[1]{\ens{\left\lbrace #1 \right\rbrace}}
\newcommand{\betrag}[1]{\ens{\left|#1\right|}}

\newcommand{\alive}[0]{\textrm{alive}}
\newcommand{\dead}[0]{\textrm{dead}}
\newcommand{\intact}[0]{\textrm{intact}}
\newcommand{\decayed}[0]{\textrm{decayed}}
\newcommand{\Schroedinger}[0]{Schr\"odinger}
\newcommand{\Erw}[1]{\ens{\langle #1 \rangle}}

\newcommand{\aut}[2]{\textsc{#1\,#2}} %Autoren, für die Bibliographie, ggf mit \mbox für die Namen
\newcommand{\autgr}[2]{\textsc{#1 and #2}} %Autorengruppe für die Bibliographie (Trennung mit "und" oder durch Komma)
\newcommand{\tit}[1]{\textit{#1}} %Titel, für die Bibliographie
\newcommand{\zeit}[1]{#1} %Name der Zeitschrift
\newcommand{\Band}[1]{\textbf{#1}} %Bandnummer oder Jahrgang der Zeitschrift
\newcommand{\Seiten}[2]{\mbox{pp. #1--#2}} %Seitenangaben: "123 -- 256" oder "123--256"
\newcommand{\aaO}[0]{\textit{ibid.}} %a.a.O., ebenda, ebd., ibid., ...

\begin{document}

\title{Evidence of entanglement, or:\\ Is Schr\"odinger's cat really entangled?}

\author{Kedar S. Ranade}
\affiliation{Institut f\"ur Quantenphysik, Universit\"at Ulm, and Center for Integrated Quantum Science and Technology
  (IQ\textsuperscript{ST}), Albert-Einstein-Allee 11, D-89081 Ulm, Deutschland (Germany)}
\author{Kaled Dechoum} %FEHLER IN REVTEX 4.1 !!! --> Doppelte Institutszugehörigkeit wird nicht angezeigt
\affiliation{Instituto de F\'isica, Universidade Federal Fluminense, 24210-340 Niter\'oi RJ, Brazil}
%\affiliation{Institut f\"ur Quantenphysik, Universit\"at Ulm, Albert-Einstein-Allee 11, D-89081 Ulm, Germany}

\date{August 14th, 2013}


\begin{abstract}
  Quantum effects and, in particular, entanglement are by now widely recognized in all areas of physics and
  related fields. However, we feel that the precise notion of entanglement---though mathe\-matically
  well-defined---still generates confusion even in the academic environment. Our aim in this article is to point
  out and clarify some aspects of the notion of entanglement, in particular with respect to observable quantities.
  We do this by illustrating common misconceptions with \Schroedinger's cat paradox,
  explain them in mathematical terms and provide examples where misunderstandings arise and how they can be solved.
\end{abstract}

\maketitle\thispagestyle{plain}

\section{Introduction}
The notion of entanglement has by now grown to one of the most prominent features in quantum mechanics
and, in particular, in quantum information theory. The power of quantum computers and the security of quantum
cryptography, which belong to the most popularized fields of quantum theory, rely on or are closely related
to the concept of entanglement.
Nevertheless, students (and researchers) often struggle with conceptually different aspects of
entanglement used in literature. On the one hand, there are foundational aspects of physics like the EPR
paradox or Bell's inequality (there are several works on these topics, e.\,g. Ref.~\onlinecite{GHSZ}). On the
other hand there appear more
\anfEngl{applied} results of entanglement theory, including quantum cryptography, quantum teleportation,
quantum computing and the like. A point which is often missed and which we want
to address here, is that in order to observe entanglement we must be specific on which quantities can actually
be measured---and we could be even more general: in quantum mechanics, we always should specify and take into account
which quantities can be measured and thus are \emph{observables}.
\par Although mathematically well-defined, there seem to exist common misconceptions about the nature of entanglement.
We shall therefore refer to entanglement in two related, but different ways, namely
\begin{itemize}
  \item entanglement as a mathematical concept and
  \item entanglement as a resource in quantum processing.
\end{itemize}
On the one hand, with entanglement as a purely mathematical concept there is nothing mysterious or arcane about it:
it is just a property of certain states on a tensor product of Hilbert spaces (i.\,e. density matrices or vectors).
The principal difficulties in physically understanding entanglement arise only by invoking the notion of (space-time)
locality, which result in peculiar features of quantum mechanics exhibited by the violation of Bell's inequality.
On the other hand, in any physical situation where we want to make use of quantum entanglement, we must, at
least in principle, be able to prove the existence of entanglement (in the mathematical sense), which implies proving
the absence of any classical explanations for the measurements observed. This also means that entanglement can only
be described in terms of measurements, or to state it more dramatically: if there is no way of measuring it, there
is no (physically usable) entanglement.
\par In this article, we shall clarify, without any attempt to claim novelty, these aspects of the notion
of entanglement. We note that this can be taken even further: if, for any reason, we are restricted to observables
% which can jointly be measured (so-called \emph{commensurable observables}, e.\,g. the total spin $\hat{\vec S}$ and its
% $z$-component $\hat{S}_z$, but not $\hat{S}_z$ and $\hat{S}_x$), there always exist a classical description of the
which can jointly be measured (they are called \emph{commensurable} and are represented by commuting operators), there
always exist a classical description of the experiment. To summarize the main statements proposed in this article:
\begin{itemize}
  \item By measuring commuting observables only, one can never prove the existence of entanglement.
  \item One needs at least one pair of non-commuting observables to be able to detect any quantum effect.
\end{itemize}
Since the notion of entanglement was introduced by \Schroedinger{} in the very same paper where he introduced
his well-known cat paradox, we take this as the starting point for our survey and start by illustrating \Schroedinger's
cat and explain why there is no evidence for entanglement there. Then, we will give the basic mathematical formalism of
quantum mechanics and entanglement. After that, we will give some illustrations and finally summarize and
conclude our results.

\section{Short history of \Schroedinger's cat}\label{Ch_Cat}
In reply to the Einstein-Podolsky-Rosen paradox of 1935,\cite{EPR} a now famous german-language article
\anfEngl{Die gegenw\"artige Situation in der Quantenmechanik} (The present situation in quantum mechanics)
appeared in three parts in the journal \emph{Die Naturwissenschaften} (Natural sciences) in the same year,
\cite{Schroedinger} where Erwin
\Schroedinger{} coined the german term \anfEngl{Verschr\"ankung}. (The english equivalent \anfEngl{entanglement} also
appeared in a paper by \Schroedinger{} at the same time.\cite{Schroedinger2}) At the end of the first part an apparent
paradox, now known as \anfEngl{\Schroedinger's cat},
was introduced. Let us, for the sake of completeness, recall this Gedankenexperiment: We put a living cat
into a box together with an apparatus which is secured from the cat: In the apparatus,
there is a small portion of a radioactive substance (say, a single nucleus), such that the probability of
a single decay within one hour from the closing the box is one half. A decay would trigger a little hammer
which destroys a little flask containing prussic acid that would immediately kill the cat. We open the box
after one hour and check, whether the cat is alive or dead and whether the nucleus is decayed or not. If we perform
the experiment several times, we find that the cat is dead half of the times and alive half of the times,
and whenever the cat is dead, the nucleus has decayed, and whenever it is alive, the nucleus is intact; this
is a classical correlation between the cat and the nucleus.

\subsection{Quantum mechanics of the cat experiment}
The usual analysis of \Schroedinger's cat is the following: The (classical) cat has two states, dead and alive,
and so has the radioactive nucleus, decayed or not. Let us write this in the so-called Dirac notation: the states
of the cat are $\ket{\alive}$ and $\ket{\dead}$, those of the nucleus $\ket{\intact}$ and $\ket{\decayed}$. The joint
system of cat and nucleus now has---by classical reasoning---four possible states, which we may write as
1. $\ket{\alive}\ket{\intact}$, 2. $\ket{\alive}\ket{\decayed}$, 3. $\ket{\dead}\ket{\intact}$ and
4. $\ket{\dead}\ket{\decayed}$. In our model---the poison kills the cat and the cat does not die of other
causes---we have correlations (in the statistical sense) between the
state of the cat and the state of the nucleus, namely that possibilities 2 and 3 are excluded. Note that up to
now there is nothing \anfEngl{quantum} here, in particular, these correlations are not entanglement!
\par Any (even popular-science) introduction to quantum mechanics tells us that we can have superpositions of states.
In particular, we could have a superposition of the possibilities 1 and 4 from above. As we do not know anything about
the system, except that the nucleus has decayed with probability $1/2$, we may be tempted to conclude that the system,
consisting of the cat and the nucleus, is in the superposition state
\begin{equation}\label{cat1}
  \ket{\Psi} = \frac{\ket{\alive}\ket{\intact} + \ket{\dead}\ket{\decayed}}{\sqrt{2}},
\end{equation}
where the normalization $\frac{1}{\sqrt{2}}$ represents the square root of the probabilities as the wave
function is the probability amplitude, and only its square root the probability density. This is---in quantum
mechanics---a perfectly valid statement; in popular science one would say that, before the measurement, the cat
is dead and alive at the same time.
If we did not wait for precisely one half-life time of the nucleus, we would get an asymmetry for the two
possibilities, since the probability of the decay or non-decay of the nucleus is not 50\% each. We could
incorporate this asymmetry, if not either probability is zero or one, but this will not significantly change our
line of argument.
\par Nevertheless, up to now our reasoning to infer the state of the system is incorrect. The first thing, which
may appear strange, is the lack of a phase. We know that quantum theory \anfEngl{lives} in a complex Hilbert space,
where we have the choice of a complex phase $\EZ^{\iE\phi}$. Global phases are irrelevant, but relative phases
are not. So, apparently, for any choice of the angle $\phi \in (0;\,2\pi)$, another equivalent state describing
the system would be
\begin{equation}\label{cat2}
  \ket{\Psi(\phi)} = \frac{\ket{\alive}\ket{\intact} + \EZ^{\iE\phi} \ket{\dead}\ket{\decayed}}{\sqrt{2}}.
\end{equation}
If we repeat the experiment several times, we will see that the cat is dead half of the time and alive
half of the time, which completely reproduces our results from above. So in terms of our observations we cannot distinguish
a state with some $\phi$ from another one with $\phi^\prime$. Now this is not a problem, since quantum
mechanics tells us that any of these states is entangled.
\par However, up to now we have dealt with pure-state quantum mechanics only. But a complete
quantum-mechanical description should also take into account the possibility of mixed states.
In eqs. (\ref{cat1}) and (\ref{cat2}) we used \emph{coherent} mixtures of quantum states, i.\,e. superpositions
of state vectors. We can also consider classical or \emph{incoherent} mixtures of states, i.\,e. convex
sums of density matrices. 
This comes in quite naturally, if we say, that we do not know the actual state of the cat---it may be any of those
mentioned in eq. (\ref{cat2})---so that we take the (statistical) \anfEngl{average} over all possible angles $\phi$.
This averaging procedure is mathematically well-defined and results in a mixed state
\begin{equation}\label{cat3}
  \rho = \frac{\pr{\alive} \x \pr{\intact} + \pr{\dead} \x \pr{\decayed}}{2},
\end{equation}
which once again exactly reproduces our observations. \emph{But this state is not entangled!} This shows that
from the description of \Schroedinger's cat experiment, we cannot conclude that the cat and
the nucleus are entangled. We do not have enough evidence of entanglement in this system.
\par At this point we have shown that just checking the $2 \times 2 = 4$ states is insufficient to prove entanglement
in a quantum state. In theoretical quantum physics, it is easy to specify a  measurement which could provide evidence
of entanglement as this is just two two-level systems, and such measurements are well-known for qubits. However, we
cannot think of a practical implementation of such measurements if a cat is to be used.

\subsection{A modified cat paradox}
To give an example of how the state in eq. (\ref{cat3}) may appear, we can slightly modify the experiment. Assume that
the nucleus exhibits a $\beta^-$-decay and emits an electron and an antineutrino. The antineutrino transcends the box
and is collected in some big neutrino detector. So the operator of that detector knows the state of the cat and may
tell the state to our experimentalist. In this case the system is composed of three subsystems: the cat, the
nucleus and the existence of the antineutrino (with two states: it exists or it does not exist). As we only consider
the two systems of cat and nucleus, we have to trace out the third system, which results in the state described in
eq.~(\ref{cat3}). If the neutrino detector did not exist, the neutrino would leave the earth undetected, but this
cannot alter the state. So the observer cannot distinguish between the states of eq.~(\ref{cat2}) and eq.~(\ref{cat3}).
\par The deeper reason for this effect is that it is not possible to prove that a quantum system is entangled
by a single measurement whatsoever (in the statistical sense, i.\,e. repeating the experiment several times).
In the example of the cat, the only measurement we perform is the dead-alive-measurement of the cat and, accordingly,
the intact-decayed-measurement of the nucleus. To phrase it even more generally: No system can be shown to exhibit
genuine quantum phenomena, if the observer is restricted to a single measurement. All quantum phenomena rely on
measurements which are non-commensurable.

\section{Mathematical formalism}
In order to present our statements in a more general way, we should introduce some formalism. We start
with a classical system, on which we can perform measurements. In our case, we use a reduced model
of the cat, i.\,e., we do not consider its position or momentum, its color, its sex or whatever
feature a cat may have. We only consider, whether it is dead or alive, and our measurement thus has
a binary outcome. We must assume that the other features are unrelated to whether the cat is dead or alive.
\par As an axiom, for a quantum system there is always a Hilbert space associated with the system.
The dimension of that space is the number of classical outcomes, which may be finite or infinite.
In case we have two classical outcomes, like in our description of the cat and the nucleus,
respectively, the Hilbert space is the two-dimensional complex Hilbert space $\cH = \C^2$. (Note that
two Hilbert spaces of the same dimension are isomorphic, and
that for every such dimension a Hilbert space can be constructed. For quantum mechanics, we have
$\cH = \C^n$, $n \in \N$, if the number of outcomes is finite. When we have a continuous measurement, like the
position, we have an infinite-dimensional space and things get much more difficult, but we shall ignore that.)
Combining two systems with $n$ and $m$ classical
outcomes (in our case $n = m = 2$), the joint system may have $n \cdot m$ outcomes, and the associated
mathematical construction is the tensor product of Hilbert spaces. In our case, the Hilbert space of
the joint system is thus four-dimensional, i.\,e. $\cH = \C^2 \x \C^2 \cong \C^4$.
\par Another axiom of quantum mechanics is that observables correspond to operators on Hilbert space with
their eigenvalues being the possible measurement outcomes. As we require that the outcome of a measurement
be real (and not imaginary or complex), these operators need to be self-adjoint. A \emph{state} now is the mathematical
object which maps to each measurable operator its expectation value as a weighted mean of all possible outcomes.
So an operator $\hat{A}$ is mapped to $\Erw{\hat A}$, and all we may know about the state is this expectation for
any measurable operator. How to specify this mapping? This can be done in terms of density matrices $\rho$,
where we have $\hat{A} \mapsto \Erw{\hat A} = \Spur(\rho\hat A)$. If (and only if) $\rho$ is a one-dimensional
projector, there is a state vector $\ket{\Psi} \in \cH$, such that $\rho = \pr{\Psi}$, and $\ket{\Psi}$ is unique
up to a phase.
Note now that we cannot directly measure the state of a quantum system. We can only measure expectation values
of operators. For two different density matrices, $\rho$ and $\rho^\prime$, there always exist an operator
$\hat{A}$, such that $\Spur(\rho\hat A) \neq \Spur(\rho^\prime\hat A)$, i.\,e., we can distinguish $\rho$ and
$\rho^\prime$ by their statistical properties, if indeed $\hat A$ represents a measurable quantity.
In the best case, we may know the complete mapping $\hat{A} \mapsto \Erw{\hat A}$ for all self-adjoint
operators $\hat A$, so that $\rho$ can be reconstructed uniquely. But if, for physical, mathematical or practical
reasons, not all self-adjoint operators $\hat{A}$ on the Hilbert space correspond to measurable quantities, it may
happen that there are two (or more) density operators, say $\rho$ and $\rho^\prime$, which yield the same result
for all measurable quantities.
%To phrase it in other words: the set of self-adjoint operators on $\cH$ is $\fS = \cB(\cH)_{\mathrm{sa}}$,
%but the set corresponding to measurable quantities is a proper subset $\fA \subset \fS$. For any mapping
Now we may take two different points of view (regarding the question of the nature of a quantum state, see also
Ref.~\onlinecite{Newton}):
\begin{itemize}
  \item The density operators $\rho$ and $\rho^\prime$ are different states, but we do not have means to
    distinguish them (e.\,g. we can measure position and momentum of an atom, but do not have a magnetic field to
    determine spin).
  \item Since all measurable quantities have the same statistical properties, $\rho$ and $\rho^\prime$
    \emph{are} actually the same physical state.
\end{itemize}
While the second interpretation is more physically motivated, we take the first, which is easier to handle in
our discussion, but the results will be the same. So for a quantum system with a certain set of measurable
quantities, to which (by assumption or measurement) we know that there are assigned certain outcomes, we find
a set of compatible density operators. In the following section we shall define the mathematical notion of
entanglement for a single density operator; for our system, the entanglement then could be the entanglement
of any of these compatible density operators. In quantum information, entanglement is used as a resource, and
it turns out that we have to minimize the entanglement over all compatible operators in order to find the
usable entanglement for physical tasks.

\subsection{Entanglement}
There is a vast amount of literature on entanglement, e.\,g. the review by the Horodeckis,\cite{Horodecki} 
and we do not want to go into too many details here. As such, \emph{entanglement is a property of Hilbert spaces
and their tensor products}. The tensor product of two Hilbert spaces $\cH_A$ and $\cH_B$ is the Hilbert space
$\cH := \cH_A \x \cH_B$ of all superpositions of vectors $\ket{\psi}_A \x \ket{\phi}_B$, where $\ket{\psi}_A \in \cH_A$
and $\ket{\phi}_B \in \cH_B$. Now, any state on such a system, described by a state vector or a density matrix, is
either \emph{separable} or not; inseparable states are called \emph{entangled}. (Note that we only consider bipartite
entanglement here, i.\,e. we only have two subsystems; otherwise the theory gets more complicated.)
The mathematical definition of entanglement is the following: A (state) vector
$\ket{\Psi}_{AB} \in \cH$ is separable, if it can be written in the form $\ket{\psi}_A \x \ket{\phi}_B$, i.\,e.,
without the need of any superposition.
\par How to define entanglement for mixed states? Any mixed state can (in general, non-uniquely) be written in the
form $\rho_{AB} = \sum_{i = 1}^{n} p_i \ket{\Psi_i}_{AB}\bra{\Psi_i}$ for some state vectors $\ket{\Psi_i}_{AB} \in \cH$
and $p_i \geq 0$ with $\sum_{i = 1}^{n} p_i = 1$. By definition, a state is separable, if there exists such a
decomposition, where all the $\ket{\Psi_i}_{AB}$ are separable. This is the definition of entanglement, which we want
to invoke, and all other concepts which appear in literature are derived from that.
\par We may also note the Schmidt decomposition (cf. e.\,g. Ref\/s.~\onlinecite{EkertKnight} or \onlinecite{NielsenChuang}):
for any vector
$\ket{\Psi}_{AB}$, there exist orthonormal systems $(\ket{\psi_i}_A)_{i = 1}^{n}$ and $(\ket{\psi_i}_B)_{i = 1}^{n}$
and $(\lambda_i)_{i = 1}^{n}$ with $\sum_{i = 1}^{n} \betrag{\lambda_i}^2 = 1$, such that
\begin{equation}
  \ket{\Psi}_{AB} = \sum_{i = 1}^{n} \lambda_i \ket{\psi_i}_A \x \ket{\psi_i}_B.
\end{equation}
The state is thus separable, if and only if only one of the $\lambda_i$ is non-zero, and more generally it is possible
to build entanglement measures based on the distribution of the $\lambda_i$. %Harshman-Ranade, Bertlmann u.a.

\subsection{State determination and tomography}
Now consider the following task: we are given a quantum state $\rho$, whether pure or not, and we have to determine
that state. (This question in itself gives rise to a large amount of problems, cf. e.\,g. Ref.~\onlinecite{ParisRehacek}.)
As we can only perform measurements and measurements, in general, alter the state, there is, by principle,
no way to determine the state. If we relax the requirements, we can move on to a related task: given $n$ copies of
identically prepared states, can we recover the state? Quantum information theory tells us that we can determine
the state to arbitrary precision, provided $n \rightarrow \infty$; this is known as state tomography. But there is
always the necessity to perform different measurements. A measurement in quantum information corresponds to a basis
of a Hilbert space and thus to possible observables. Such a measurement will thus only give us information on the
diagonal elements of a density matrix in that basis, but not on the off-diagonal elements. We will explain in the
following that true quantum effects can be observed, if and only if at least two different bases are used for
measurements or, in other words, at least two observables do not commute.

\section{Applications and examples}
In this section we use the mathematical formalism to analyze the cat paradox and other applications in physical
situations.

\subsection{Cat-nucleus system}
We start by examining \Schroedinger's cat. The joint system of the cat and the nucleus has the four classical
states of section \ref{Ch_Cat}. It is therefore described by a $4 \times 4$-density matrix, whose rows and
columns we order as before: 1.~$\ket{\alive}\ket{\intact}$, 2. $\ket{\alive}\ket{\decayed}$,
3. $\ket{\dead}\ket{\intact}$ and 4. $\ket{\dead}\ket{\decayed}$. From our repeated measurements we can try to infer the
density matrix of the cat-nucleus system:
\begin{equation}
  \rho = \begin{pmatrix}
           a_{11} & a_{12} & a_{13} & a_{14}\\
           a_{21} & a_{22} & a_{23} & a_{24}\\
           a_{31} & a_{32} & a_{33} & a_{34}\\
           a_{41} & a_{42} & a_{43} & a_{44}\\
         \end{pmatrix}.
\end{equation}
Which observables do we have? We have the cat dead/alive-observable $\sigma_z^{(1)}$ and the nucleus
decayed/intact-observable $\sigma_z^{(2)}$ resulting in the joint observable $\sigma_z^{(1)} \x \sigma_z^{(2)}$
with four possible outcomes.
The probabilities that the four possibilities appear, are now $a_{11},\,a_{22},\,a_{33}$ and $a_{44}$. From
our observations we know $a_{11} = a_{44} = 1/2$ and $a_{22} = a_{33} = 0$.
\par Now the density matrix must be positive semidefinite and, in particular, hermitian. Positive semidefiniteness
can be checked by the Hurwitz-Sylvester (principal minor) criterion:\cite{Gantmacher} an $m \times m$ matrix is positive
semidefinite, if and only if all its principal minors are non-negative. The principal minors are the determinants
of the $2^m-1$ submatrices (including the matrix itself), where a set of corresponding rows and columns of
the original matrix is left out. In our case this immediately implies that all non-diagonal elements except
$a_{14}$ and $a_{41}$ must vanish. By hermiticity, we are left with a parameter $a_{41} = a_{41}^* =: x$.
\par From our observations we have found the complete matrix up to a single complex parameter which must fulfill
$a_{11}a_{44}-a_{14}a_{41} = 1/4 - \betrag{x}^2 \geq 0$ or $\betrag{x} < 1/2$ for positivity. But our measurement
by no means reveals $x$. How to check whether $\rho$ is entangled? We make use of the Peres-Horodecki criterion
for $2 \times 2$ systems\cite{Peres,HorodeckiPPT} stating that the partial transpose of a density matrix---the
transpose with respect to either subsystem---is positive semidefinite, if and only if the state is separable.
We calculate the non-zero eigenvalues of the partially transposed density matrix and find $+\betrag{x}$ and
$-\betrag{x}$. So our observations are consistent with no entanglement $x = 0$, maximal entanglement $\betrag{x} = 1$
and any amount of partial entanglement $\betrag{x} \in (0;1)$.
\par To conclude, we can only determine the diagonal elements of the density matrix. But the quantum properties and,
in particular, the entanglement of any system are related to the off-diagonal elements of the density matrix.
\emph{We have learned nothing about the entanglement of the system!}
It turns out that it order to have evidence for entanglement and, more generally, any \anfEngl{quantum}
property as superposition, we must have at least two non-commuting observables. Moreover, this is also
sufficient as the examples of Bell inequalities or quantum cryptography show, since in this case, we can always
find off-diagonal values which indicate true quantum behavior.

\subsection{Quantum cryptography}
The cat-nucleus system may seem somewhat artificial. Now we want to point out that
evidence of entanglement is of practical value in quantum information theory. One of the applications of quantum
technology is quantum cryptography. In quantum cryptography two distant parties, Alice and Bob, want to share a message
in such a way that an eavesdropper, Eve, cannot infer the content of the message. The use of the one-time pad enables
them to do so, provided they share a key, i.\,e., a large enough random number Eve does not know. Suppose Alice
and Bob share two qubits (e.\,g. photons with their polarization) in the pure entangled state\cite{Ekert}
\begin{equation}\label{PhiPlus}
  \ket{\Phi^+}_{AB} = \frac{1}{\sqrt{2}}\left[\ket{0}_A\ket{0}_B + \ket{1}_A\ket{1}_B\right],
\end{equation}
which is sufficient to prove security. If Alice and Bob measure that state in the basis $\Mg{\ket{0},\,\ket{1}}$,
they get with equal probability as outcome both zero or both one, i.\,e. a random bit. Since the state is pure,
it cannot have any correlations with other systems, in particular, not with Eve's system. But only from measurement
in the basis $\Mg{\ket{0},\,\ket{1}}$, Alice and Bob cannot say that Eve does not know the bit. Consider the
alternative three-qubit state, the GHZ state
\begin{equation}\label{PhiGHZ}
  \ket{\Phi_{\text{GHZ}}}_{ABB^\prime}
    = \frac{1}{\sqrt{2}}\left[\ket{0}_A\ket{0}_B\ket{0}_{B^\prime} + \ket{1}_A\ket{1}_B\ket{1}_{B^\prime}\right]
\end{equation}
with Eve possessing $B^\prime$.
If Eve measures her state in the $\Mg{\ket{0}_{B^\prime},\,\ket{1}_{B^\prime}}$ basis she exactly knows what the random bit is,
and this state obviously cannot be used for quantum cryptography. What is the difference between these two states? Alice
and Bob would have to determine their reduced state, which in the latter case turns out to be the separable mixed state
% \begin{equation}
%   \rho_{AB} = \frac{\pr{0} \x \pr{0} + \pr{1} \x \pr{1}}{2}.
% \end{equation}
\begin{equation}
  \rho_{AB} = \frac{\ket{0}_A\bra{0} \x \ket{0}_B\bra{0} + \ket{1}_A\bra{1} \x \ket{1}_B\bra{1}}{2}.
\end{equation}
As the reader may have noted by now, this is mathematically completely equivalent to \Schroedinger's cat and its
modification (sec.~\ref{Ch_Cat}), and thus all conclusions apply here. So in order to be secure the parties in
quantum cryptography need to check their entanglement, and this is where the measurement of non-commuting operators
appear in all quantum cryptographic protocols.
\par One may ask how such ambiguity between two states appears in a real experiment. In a theoretical description
of quantum cryptography, Alice prepares the two-photon state of eq.~(\ref{PhiPlus}) and immediately measures her
part in order to prepare the single-photon state to be sent to Bob. However, in a quantum-optical setting, a weak
laser pulse is commonly used to produce photons. If the photon number of the laser is Poisson-distributed, by lowering
the average number of photons, the fraction of pulses containing two or more photons may go down, but never reaches
zero. The case of two photons is then theoretically described by the state of eq.~(\ref{PhiGHZ}), where photons
$B$ and $B^\prime$ are sent to Bob. But then Eve may just keep the photon of $B^\prime$; while Bob will correctly
measure photon $B$ (but does not measure the number of photons), Eve may infer his outcome from photon $B^\prime$.
(In quantum cryptography, this is known as the photon-number splitting attack.\cite{PNS-Luetkenhaus})

\subsection{Entanglement with the vacuum}
Another question, which is sometimes raised, is the following: consider a single photon which crosses a 50:50
beam splitter into two channels, namely the reflected one (channel 1) and the transmitted
(channel 2). So the input field is $\ket{1}$ and classically the output fields are either $\ket{0}\ket{1}$
(transmission) or $\ket{1}\ket{0}$ (reflection). If the transformation is unitary, the state after passing
the beam splitter is something like $\frac{\ket{0}\ket{1} + \ket{0}\ket{1}}{\sqrt{2}}$. But how can it be that
the vacuum, \anfEngl{nothing}, is entangled? This question is resolved by the following. First we have to distinguish
the vacuum state $\ket{0}$ from the Hilbert space \anfEngl{nothing}, the nullvector. So vacuum means \anfEngl{no
photon}, but not \anfEngl{no state}. Even if it is not populated the mode exists, and it is the entanglement between
the states of the mode. (For further details on such questions, cf. e.\,g. Ref.~\onlinecite{vanEnk}.)
\par To phrase it differently: Consider two harmonic oscillators (two modes). On the one hand, we may consider
the ground state, which in the Fock basis is written as $\ket{0}$ and interpreted as \anfEngl{no photons}. However,
this state obviously has a non-vanishing wavefunction, and a change of basis should not affect a physical quantity
like entanglement. It is not the state which is entangled, but the system in a specific state.

\section{Summary}
In this article we have worked out what is needed for a state so that it can be considered to be entangled
and illustrated this by \Schroedinger's cat paradox. We pointed out that the physically usable entanglement
must take into account the possible observables and that in order to observe quantum phenomena we need at
least two non-commuting observables. The examples we gave can always be resolved by a careful inspection of
the things we look at: what is the system, how is it (and its measurements) described in Hilbert space and
which information we can mathematically infer by measurements? This is all what is needed to evade
%a large number of
seemingly strange observations.
\par Finally we wish to remark that we should always consider density
operators, since---to modify \Schroedinger's statement---it is the density matrix (and not the wavefunction)
which is the catalog of expectations.

\section*{Acknowledgments}
The authors thank Nathan Harshman (American University, Washington D.\,C.) for several helpful comments.
K.\,S.\,Ranade acknowledges financial support by BMBF/QuOReP.

\end{document}